\documentclass[12pt]{article}
\usepackage{epsfig}
\usepackage{amsfonts}
\usepackage{latexsym}
\usepackage{amsmath,amssymb}
\usepackage{verbatim}
\usepackage{mathtools}
\usepackage{youngtab}
\def\half{{1\over 2}}
\numberwithin{equation}{section}

\def\e{{\epsilon}}

 \def\p{\partial}

 \def\r{\rightarrow}

\def\s{\sigma}

\newcommand{\bea}{\begin{eqnarray}}
\newcommand{\eea}{\end{eqnarray}}
\newcommand{\be}{\begin{equation}}
\newcommand{\ee}{\end{equation}}
\newcommand{\ba}{\begin{align}}
\newcommand{\ea}{\end{align}}
\newcommand{\bmat}{\begin{bmatrix}}
\newcommand{\emat}{\end{bmatrix}}

\renewcommand{\L}{\Lambda}
\newcommand{\bigO}{\mathcal{O}}

\newcommand{\W}{\mathcal{W}}
\newcommand{\C}{\mathcal{C}}
\newcommand{\tr}{\mbox{tr}}

\newcommand{\CC}{\mathbb{C}} 
\newcommand{\RR}{\mathbb{R}} 
\newcommand{\ZZ}{\mathbb{Z}} 

\newcommand{\M}{\mathcal{M}}

  \makeatletter
  \let\over=\@@over \let\overwithdelims=\@@overwithdelims
  \let\atop=\@@atop \let\atopwithdelims=\@@atopwithdelims
  \let\above=\@@above \let\abovewithdelims=\@@abovewithdelims

\DeclareMathOperator{\diag}{diag}

\def\a{\alpha}

\def\d{\delta}
\def\e{\epsilon}

\def\l{\lambda}
\def\L{\Lambda}

\def\p{\pi}
\def\r{\rho}

\def\s{\sigma}

\def\half{{1 \over 2}}
\def\nonu{\nonumber \\{}}

\begin{document}

\begin{titlepage}
\begin{flushright}
NSF-KITP-11-222
\end{flushright}
\vskip 2cm
\begin{center}
{\LARGE Conical Defects in Higher Spin Theories}

\vspace{6mm}
{Alejandra Castro$^a$, Rajesh Gopakumar$^b$, Michael Gutperle$^c$ and Joris Raeymaekers$^d$}\vspace{6.0mm}\\
\bigskip
\centerline{$^a$ {\it McGill Physics Department, 3600 rue University, Montreal, QC H3A 2T8, Canada} }
\medskip
\centerline{$^b$ {\it Harish-Chandra Research Institute, Chhatnag Rd., Jhusi, Allahabad 211019, India}}
\medskip
\centerline{$^c$ {\it Department of Physics and Astronomy, UCLA, Los Angeles, CA 90095, USA}}
\medskip
\centerline{$^d$ {\it Institute of Physics of the ASCR,
Na Slovance 2, 182 21 Prague 8, Czech Republic}}

\vspace{1cm}

\begin{abstract}
\noindent
We study conical defect geometries in the $SL(N)$ Chern-Simons formulation of higher spin gauge theories in AdS$_3$.  We argue that (for $N\geq 4$) there are special values of the deficit angle for which these geometries are actually smooth configurations of the underlying theory.
We also exhibit a gauge in which these geometries can be viewed as wormholes interpolating between two distinct asymptotically AdS$_3$ spacetimes. Remarkably, the spectrum of smooth $SL(N, \CC)$ solutions, after an appropriate analytic continuation, exactly matches that of the so-called ``light primaries'' in the minimal model ${\cal W}_N$ CFTs at finite $N$. This gives a candidate bulk interpretation of the latter states in the holographic duality proposed in \cite{Gaberdiel:2010pz}.

\end{abstract}

\end{center}

\setcounter{footnote}{0}

\end{titlepage}
\newpage

\baselineskip 17pt

\tableofcontents


\section{Introduction}

The work of Vasiliev and collaborators\footnote{See  \cite{Bekaert:2005vh} for a review and a comprehensive list of references.} gives a classical description of gravity in AdS spacetimes consistently interacting with an infinite set of massless higher spin gauge fields.  These theories are interesting since  they possess some features of full fledged string theories. They include an infinite tower of higher spin fields (though a much smaller infinity than in string theory) with interactions involving an infinite number of derivatives, and a much larger gauge invariance than diffeomorphism invariance. On the other hand, one can write down background independent classical equations of motion, and the structure of these equations  is considerably simpler than those of classical string (field) theory. These theories have  attracted a lot of attention from the perspective of the AdS/CFT correspondence. One may view them as either describing a sub sector of the string theory dual to free Yang-Mills theory or as duals to vector like field theories (see e.g.  \cite{Konstein:2000bi,Sundborg:2000wp,Mikhailov:2002bp,Sezgin:2002rt,Klebanov:2002ja,Giombi:2009wh,Giombi:2010vg,Gaberdiel:2010pz, Douglas:2010rc}).

Apart from providing new examples of AdS/CFT dualities, higher spin theories can also be useful toy models of stringy gravity. Specifically, one may test whether properties of (semi-)classical gravity minimally coupled to matter, such as the existence of horizons and curvature singularities, are affected by the nonlinear and nonlocal coupling of an infinite tower of higher spin fields to gravity.

The three dimensional Vasiliev theories are an especially tractable arena since, like three dimensional gravity, these theories do not have any propagating degrees of freedom for the higher spin fields and yet possess interesting classical solutions like black holes.  At the same time, there is a conjecture \cite{Gaberdiel:2010pz} relating a particular three dimensional Vasiliev theory (based on the higher spin algebra $hs[\lambda]$ and coupled to two additional complex scalars) to the large N 't Hooft limit of the
$\W_N$ coset CFT
\be\label{coset}
{SU(N)_k \otimes SU(N)_1 \over SU(N)_{k+1}}~.
\ee
The 't Hooft limit consists of taking $N,k\to \infty$ while keeping $\lambda= N/(k+N)$ fixed.
This gives a candidate microscopic description for an interesting
class of three dimensional higher spin theories in which one can address some of the above issues.
For further developments on this conjecture, including generalizations in various directions, see \cite{Gaberdiel:2011wb,Gaberdiel:2011zw,Chang:2011mz, Ahn:2011pv, Gaberdiel:2011nt, Papadodimas:2011pf, Ahn:2011by, Ouyang:2011fs, Castro:2011zq, Creutzig:2011fe}.
See also \cite{Castro:2010ce, Castro:2011ui} for a discussion on black holes and the stringy exclusion principle in these theories.

In some ways the simplest  class of three dimensional higher spin theories are those that can be  formulated as Chern-Simons theories based on the gauge group $SL(N,\RR)\times SL(N,\RR)$ \cite{Blencowe:1988gj}. These theories have (in addition to gravity) only a finite number of higher spin fields  of spin $s=3,4,\cdots, N$ and no additional scalars.  They also have an enlarged ${\cal W}_N$ asymptotic symmetry \cite{Henneaux:2010xg, Campoleoni:2010zq, Gaberdiel:2010ar, Gaberdiel:2011wb, Campoleoni:2011hg}. In \cite{Gutperle:2011kf,Ammon:2011nk,Kraus:2011ds,Castro:2011fm, Tan:2011tj} black hole solutions in these theories which carry higher spin charge were constructed. It was found that the notion of event horizon or even its existence is a gauge dependent statement. An invariant characterisation is via holomomy conditions for the Chern-Simons connection along the Euclidean time circle. It was shown that imposing appropriate conditions on this holonomy is equivalent to the existence of a first law of thermodynamics. Further, in \cite{Kraus:2011ds}, the resulting thermodynamic
partition function in the bulk (with chemical potential for the spin-3 charge) was successfully compared with the prediction from the large $N$ CFTs (\ref{coset}) at $\lambda=0,1$.

In the  present paper, we  study the nature and the possible resolution of conical singularities\footnote{There is a large literature on conical singularities in three dimensional gravity starting with \cite{Deser:1983tn,Deser:1983nh}.} in the $SL(N, \RR) \times SL(N, \RR)$ Chern-Simons formulation of higher spin gauge theories in AdS$_3$, and their role in the duality of \cite{Gaberdiel:2010pz}. For $N \geq  4$, we argue that there are special values of the deficit angle for which these geometries are actually smooth configurations of the underlying theory. This follows from the fact that the eigenvalues of the holonomies of the $SL(N, \RR) \times SL(N, \RR)$ connection along the contractible spatial circle are gauge invariant. A gauge invariant characterisation  of smoothness is that the holonomy matrix be a trivial element of the gauge group. This condition is satisfied only for a discrete set of configurations. We thus find a discrete spectrum of states whose number grows rapidly as we increase $N$. In fact, they form a densely spaced discretuum in the large $N$ limit.

The Lorentzian theory also admits smooth configurations which correspond to a conical {\it surplus}.
At first sight, one might reject these solutions on the ground that they fall below the vacuum energy of AdS and thus seem to correspond to negative energy excitations. However, we give evidence that these solutions have an important role to play in relation to the duality proposed in \cite{Gaberdiel:2010pz}. In fact, we propose an analytic continuation of the parameters in the coset (\ref{coset}) which relates it to the $SL(N, \RR)\times SL(N, \RR)$ gravitational theory which is the focus of our study here.

Specifically, in (\ref{coset}), we keep $N$ fixed and take $k \rightarrow -(N+1)$.\footnote{This analytic continuation takes $\lambda \rightarrow -N$ which is believed to be the analytic continuation between the $hs[\lambda]$ theory and the $SL(N,\RR)$ theory \cite{Feigin88, Fradkin:1990qk, Gaberdiel:2011zw}.} Under this analytic continuation we find that a conical surplus carries identical charges as a primary in the coset labelled by representations $(\Lambda ,\Lambda)$   for any finite $N$. Here $\Lambda$ is a representation of $SU(N)$. Indeed, after analytic continuation the spectrum of surpluses has the right sign, i.e. of excitations above the CFT vacuum.
Recall that in the large $N$ limit, these primaries became light \cite{Gaberdiel:2010pz, Gaberdiel:2011zw, Papadodimas:2011pf} and formed a discretuum above the vacuum, which is exactly the behavior of the surpluses (as for the deficits).
However, while the conical surpluses in the Lorentzian theory match with some of the light $(\Lambda, \Lambda)$ primaries, not all of the latter can be interpreted this way.

Surprisingly, the analogous spectrum of smooth connections in the higher spin theory with Euclidean signature (i.e. based on the gauge group $SL(N, \CC)$) seems to exactly account for {\it all} the light primaries.
Revisiting the smoothness constraint on the connections in the $SL(N,\CC)$ Chern-Simons theory we find that the theory posseses significantly more smooth solutions in Euclidean signature. This is due to both the presence of a $\ZZ_N$ center in  $SL(N,\CC)$, and the absence of a reality constraint on the connection.
These additional states do not have a natural interpretation in the Lorentzian theory. Even though the metric in these cases is real in both Euclidean and Lorentzian signature, it is the higher spin fields which cease to be real after a Wick rotation. Understanding in detail the Wick rotation  for higher spin theories, and its implications for the spectrum of the theory, is an interesting question that we leave for future work.

In any case, we find the spectrum of dimensions of the smooth configurations in the $SL(N, \CC)$ theory  identically agrees  with the analytic continuation of the dimensions of the light primaries in the ${\cal W}_N$ minimal models, for any finite $N$. Moreover, we also perform a very nontrivial match of the spin three and spin four charges on both sides.
We take this to be strong evidence that the conical surpluses -- and other smooth configurations --  can be appropriately identified as the holographic dual to these primaries of the ${\cal W}_N$ CFT. We may equivalently take this as evidence that the
$hs[\lambda]$ theory contains smooth solutions whose spectrum is the same as that of the $(\Lambda, \Lambda)$ primaries. Even though it is challenging, a direct construction of these configurations would provide a robust definition of these physical states in the $hs[\lambda]$ theory.

The organization of the paper is as follows. We briefly review, in section \ref{sec:hol}, the Chern-Simons formulation
of higher spin theories in AdS$_3$ as well as our criterion for smoothness of configurations in this theory. Section \ref{sec:CD} is devoted to constructing the conical defect solutions in the gauge theory and identifying the ones which satisfy the smoothness criterion. We find the discrete spectrum of conical deficits whose energies lie above the global AdS vacuum. We mainly work in the so-called principal embedding of the gravitational $SL(2,\RR) \times SL(2, \RR)$ in the $SL(N, \RR) \times SL(N, \RR)$ gauge group and only briefly consider other embeddings. In section \ref{sec:CSur} we generalize the analysis to conical surpluses and the conditions under which they obey the standard Drinfeld-Sokolov  reduction. We also discuss the new states which appear as smooth configurations in the Euclidean $SL(N,\CC)$ theory, leading to a larger discrete spectrum. Finally in section \ref{sec:WN}, we make a detailed match of the above bulk states to the $(\Lambda, \Lambda)$ primaries in the ${\cal W}_N$ minimal models after analytic continuation. We also check the nontrivial agreement of the spin three and spin four charges on both sides. A couple of appendices carry our conventions and details of normalization.

\section{Chern-Simons formulation of higher spin theories}\label{sec:hol}

It is well known \cite{Achucarro:1987vz,Witten:1988hc} that in three dimensions Einstein gravity with a negative cosmological constant can be reformulated  as a $SL(2,\RR)\times SL(2,\RR)$ Chern-Simons theory.  Generalizing the gauge group to $SL(N,\RR) \times SL(N,\RR)$  produces gravity coupled to symmetric tensors of spin $3,4,\cdots N$
\cite{Blencowe:1988gj}. In the following  we will review some essential features of these theories which are important for our discussion.

The action of the  $SL(N,\RR) \times SL(N,\RR)$   Chern-Simons theory is given
by\footnote{Here we are assuming that the theory is parity invariant. If we allow the
levels of the two Chern-Simons factors to be different one finds a higher spin analogue of topologically massive gravity. For a study of these theories at the semi-classical level see \cite{Chen:2011vp, Bagchi:2011vr, Bagchi:2011td, Chen:2011yx}.}
\be\label{actiona}
S=S_{CS}[A]-S_{CS}[\bar A]~,
\ee
where
\be\label{actionb}
S_{CS}[A]={k_{\rm cs}\over 4\pi}  \, \tr \int_{\M} \Big( A\wedge dA+ {2\over 3} A\wedge A\wedge A\Big)~.
\ee
Here the trace `$\tr$' denotes the invariant quadratic form of the Lie algebra, and $\M$ is the 3-manifold that supports the $SL(N,\RR)$ connections $A$ and $\bar A$. The equations of motion following from (\ref{actiona})
\be
dA+ A\wedge A =0~, \quad d\bar A+ \bar A\wedge \bar A =0~,
\ee
are nothing but the flatness conditions on the connections.

The metric and higher spin fields are obtained from the Chern-Simons connection as follows:  in line with the pure gravity case, one defines  a $SL(N,\RR)$ valued generalized vielbein and spin connection
\be\label{vielbeina}
e={\ell_{A}\over 2} \big(A-\bar A\big)~, \quad \omega= {1\over 2}\big( A+\bar A\big)~,
\ee
where we introduced the  AdS radius $\ell_{A}$. The metric and higher spin fields can then be expressed in terms of trace invariants of the vielbein \cite{Campoleoni:2010zq,Campoleoni:2011hg}. For example, the metric and the spin three field can be expressed -- up to an overall constant $\epsilon_N^{(s)}$ -- as follows
\be\label{metric}
g_{\mu\nu} ={1\over \epsilon^{(2)}_N}  \tr\big (e_\mu e_\nu\big)~ , \quad  \phi_{\mu \nu\rho} = {1\over \epsilon^{(3)}_N} \tr \big(e_{(\mu } e_{\nu} e_{\rho)}\big)~,
\ee
and analogous expressions for the remaining fields. The gauge transformations of the Chern-Simons theory are mapped to generalized frame rotations as well as diffeomorphisms of the metric and  gauge transformations of the higher spin fields.  An important feature  of this construction, as emphasized in  \cite{Campoleoni:2010zq},  is that the higher spin gauge transformations act nontrivially on the metric. Consequently, properties of spacetimes such as the causal structure and the presence of curvature singularities which are coordinate invariants in theories of pure gravity  become gauge dependent in the Chern-Simons formulation of higher spin gravity.

The definitions \eqref{vielbeina} and \eqref{metric} for metric-like fields are appropriate for the principal embedding of $SL(2,\RR)$ in $SL(N,\RR)$. For this embedding the remaining generators are grouped in symmetric, traceless tensors that transform in the spin $s$ representation of $SL(2,\RR)$ which justifies formulas like \eqref{metric}.
The embedding of $SL(2,\RR)$ in $SL(N,\RR)$ is crucial to give a gravitational interpretation of the Chern-Simons theory and hence identify the matter content. Not too surprisingly,  each inequivalent embedding generates a different spectrum of the theory, and from the point of view of the Chern-Simons theory there is no preferred embedding of the gauge group. The principal embedding is popular because it is a natural and simple construction of a theory that couples gravity to fields of spin $s>2$.

This point motivates the introduction of $\ell_A$ in \eqref{vielbeina}. At this stage it looks artificial to introduce a scale given that \eqref{actionb} has no dimensionful coupling. The incentive is that the radius $\ell_A$, in conjunction with Newton's constant $G$,  will allow us to quantify the energy spectrum and compare inequivalent embeddings.  The relation between the Chern-Simons level and the gravitational couplings is
\be
k_{\rm cs}={\ell_{A} \over8 G \epsilon_N}~,
\ee
in accordance with the pure gravity limit. The central charge of the asymptotic symmetry group is \cite{Henneaux:2010xg,Campoleoni:2010zq}
\be\label{cc}
c=12k_{\rm cs} \epsilon_N = {3\ell_A\over 2 G}~.
\ee
The constant $\epsilon_N$ is defined as the normalization of the Lie algebra metric
\be\label{b:b}
\tr( L_a L_b) =\epsilon_N \eta_{a b} ~,
\ee
where $\{L_0, L_{\pm1}\}$ are the $SL(N,\RR)$ generators which form a $SL(2,\RR)$ subgroup. From \eqref{b:b} the appropriate normalization in \eqref{metric} is to set $\epsilon^{(2)}_N=\epsilon_N$.

 For a fixed representation of the gauge group, the distinct $SL(2,\RR)$ subgroups will have relative different values of $\epsilon_N$. This rescaling in particular implies that for fixed $k_{\rm cs}$, the central charge \eqref{cc} changes by a factor of $\epsilon_N$. In addition, it is clear from \eqref{metric} that the relative `size' of the spacetime changes:  the AdS radius is rescaled by a factor of $\sqrt{\epsilon_N}$. For example, in the fundamental representation of $SL(N,\RR)$ and for the principal embedding the normalization is\footnote{Explicit matrix forms of the generators of $SL(N,\RR)$ and other conventions are presented in appendix \ref{app:slN}.}
\be\label{b:bc}
\epsilon_N={1\over 12} N(N^2-1)~.
\ee

\subsection{Classification of solutions via holonomies}\label{classi}

A theory characterized by the action \eqref{actiona}  is independent of the metric on the 3-manifold $\M$, which implies that gauge invariant observables are topological invariants. A natural class are the (eigenvalues of)  holonomy matrices of the $SL(N,\RR)$ gauge fields along a closed curve. Given a closed curve $\gamma$ in $\M$,  the holonomy is defined as
\bea\label{a:b}
{\rm Hol}_\gamma (A)&=&{\cal P}\exp(\oint_{\gamma}  A)~.
\eea
Note that since we are working with real $SL(N,\RR)$ gauge fields and since the holonomy is an element of the gauge group, we do not have an $i$ in the exponent.
Traces of the holonomy in various representations are gauge independent, or equivalently, the eigenvalues of \eqref{a:b} are well-defined physical observables. Further, since the solutions of the classical equations are flat gauge connections, this operator allows us to classify all solutions in a gauge invariant manner. The Jordan decomposition of \eqref{a:b} specifies a conjugacy class, and each class labels a distinct solution of the theory.

We now proceed to describe the solutions and characterize them via the holonomy. First, we need to specify some properties of  the 3-manifold $\M$. We will assume that  the manifold is topologically $D^2 \times R$ (i.e. the same topology as global AdS$_3$). The $S^1$ of the disk $D^2$ is a contractible cycle described by $\phi\sim \phi+2\pi$.\footnote{The assumption that the $S^1$ circle is contractible immediately eliminates from the discussion black hole solutions. For a black hole the contractible cycle is time-like and the $\phi$-cycle is non-contractible; this defines the horizon of the solution. Below we will elaborate more on this point.} We will use coordinates $(\rho,\phi, t)$ and $x^\pm =t\pm \phi$.

As mentioned, the solutions to the Chern-Simons equations of motion are flat connections.  We  will, in what follows, consider connections of the form
\be\label{a:a}
A= b^{-1}a \,b +b^{-1}db \, , \, \, \qquad \bar{A}= b\bar{a} \,b^{-1} +b\, db^{-1}~.
\ee
These are flat connections only if $a$ and $\bar{a}$ have vanishing field strength.  We will consider $a=a_+dx^+$ (and correspondingly $\bar{a}=a_- dx^-$ for the connection  $\bar {A}$) with $a_+$ being a constant $SL(N, \RR)$ matrix.
We will also choose $b$ to be single valued around an appropriate contractible cycle. In fact, we will work in a  gauge where $b$ is only a function of the radial coordinate $\rho$, and  the radial component of the connection is then fixed to be $A_\rho=b^{-1}\partial_\rho b$.

For connections of the form \eqref{a:a} we have, for the holonomies along a circle of constant $\rho$ and $t$,
\bea\label{a:c}
{\rm Hol}_\phi (A)&=&{\cal P}\exp(\oint_{S^1}  A)\cr
&=&b^{-1}\exp(2 \p a_+ )\, b~,
\eea
and the analogous expression for $\bar A$.

Let us now address the question of what the admissible (or smooth) classical solutions of the higher spin theory are.
The geometric notion of smoothness is somewhat subtle in a higher spin theory since the usual curvature invariants (which one uses to characterize smoothness) are actually not  invariant under higher spin gauge transformations. However, in the present
case the higher spin gauge fields are simply $SL(N, \RR)$ gauge fields and we can use our experience from gauge theory to rephrase the question. It is natural therefore to take the criterion to be that the gauge field configuration should not be singular.

Since the classical solutions in our case have vanishing field strengths, there are no local gauge invariant observables.  The only available diagnostic are the holonomies which we mentioned earlier. Indeed if the holonomy along a contractible curve is not trivial (i.e. is not gauge equivalent to the identity element) the gauge connection must be singular somewhere in the interior of the curve. Thus we are led to requiring that the holonomy in \eqref{a:c} be the identity matrix for any admissible (smooth) classical solution.\footnote{Actually, this is the statement for $N$ odd. For $N$ even, we need to be more careful. The actual gauge group in this case is actually $(SL(N,\RR)/{\mathbb Z}_2)\times (SL(N, \RR)/{\mathbb Z}_2)$ where ${\mathbb Z}_2$ is the centre of $SL(N,\RR)$. This is because all the fields in the theory transform under the adjoint representation on which the centre acts trivially. Thus in the even case, the element which is the negative of the identity matrix is also a trivial element of the gauge group. As we will see, the holonomy of global AdS is indeed minus the identity matrix.} One immediate consequence of this requirement is that the
matrices $a_\pm$ must be diagonalizable.\footnote{A general matrix can be written as a sum of a diagonalizable matrix $D$ and a nilpotent matrix $N$ (i.e. $N^k= 0$ fore some $k$) such that
$[D,N]=0$. It is then easy to see that, if the matrix exponentiates to a multiple of the identity, the nilpotent part $N$ has to vanish.}

Note that we are considering only topologies which are the same as global AdS which led to the space-like cycle along $\phi$ being contractible. If instead the contractible cycle were time-like, such geometries would potentially correspond to black holes where the size of the cycle shrinks to zero at the horizon. In Euclidean signature it is natural to require that a solution with a well-defined and regular horizon has a single valued connection along the thermal cycle. This is the regularity condition imposed in \cite{Gutperle:2011kf,Ammon:2011nk,Kraus:2011ds} to give a gauge invariant definition of a black hole.

\section{Conical defects in the $SL(N,\RR)\times SL(N,\RR)$  theory}\label{sec:CD}

Our focus in this section will be on constructing a class of smooth solutions in the higher spin theories whose metric is locally AdS, but globally has a conical deficit. As per the discussion in the previous section, if such solutions exist these would be characterized by a trivial holonomy along the contractible $\phi$ direction. Nevertheless, their gauge field configurations would have to be distinct from that of global AdS$_3$. We will find (for $N\geq 4$) a discrete class of novel gauge configurations with this property corresponding to special values of the conical deficit angle, i.e. the mass. We should note that, in general, our solutions will also carry various higher spin charges.

\subsection{Global AdS}\label{sec:AdS}

Let us start by describing the properties of global AdS in the Chern-Simons language when the gauge group is $SL(N,\RR)$. In an appropriate coordinate system, the metric is
\be\label{b:a}
{ds^2\over \ell_{A}^2}= d\rho^2-\left(e^{\rho}+{1\over 4}e^{-\rho}\right)^2dt^2+\left(e^{\rho}-{1\over 4}e^{-\rho}\right)^2d\phi^2~.
\ee
From here it is clear that this metric is smooth (has no conical deficit) at $\rho=-\ln(2)$ where the cycle $\phi\sim \phi +2\pi$ shrinks to zero size.
The $SL(N,\RR)$ Chern-Simons connections that support this geometry are given by
\bea\label{b:aa}
A_{\rm AdS}&=& b^{-1}\left(L_{1}+{1\over 4}L_{-1}\right) b\,dx^+  + b^{-1}d\,b~,\cr
\bar A_{\rm AdS}&=&-b\left(L_{-1}+{1\over 4}L_{1}\right)b^{-1}\, dx^-  + b\,d\,b^{-1}~,
\eea
with $b=\exp(L_0\rho)$ and $x^{\pm}=t\pm\phi$.
Using the form for $L_{\pm 1}$ given in \eqref{ellone}, \eqref{ellminus} we can diagonalize the connections and find that the  holonomy around the contractible cycle $\phi\sim\phi+2\pi$ of  $A_{\rm AdS}$ is equivalent to
\be\label{b:c}
{\rm Hol}_\phi (A_{\rm AdS})\sim \exp( 2\pi \lambda_{\rm AdS}) ~,\quad
\ee
where the matrix of eigenvalues is
\be\label{b:ca}
\lambda_{\rm AdS}=\bmat \lambda_1  & 0 & \cdots &  &0 \\ 0 & \lambda_2 & 0 & \cdots \\ \vdots & &\ddots & &  \vdots \\ 0 & \cdots &  & &   \lambda_N \emat
\ee
and
\be\label{b:da}
\lambda_j= \,{i\over 2}(N+1-2j)~,\quad j=1,\ldots, N~.
\ee

According to our definition of smoothness, i.e. that the group element exponentiates to an element in the centre of the gauge group, it is clear that global AdS always corresponds to a solution with trivial holonomy. We note that for odd $N$ we get the identity and for even $N$ minus the identity. A special feature of the solution is that the eigenvalues in  \eqref{b:da}  form  an ascending sequence of distinct (half) integers.

The simple observation that we will now make is that there are flat connections other than \eqref{b:aa} which also exponentiate to a trivial element. Interestingly, as we will now show, some of these smooth connections have an appealing interpretation: in an appropriate choice of gauge, they correspond to conical defect geometries.

\subsection{When is a conical defect smooth?}\label{smoothdef}

In this section we will reverse engineer the logic by first constructing flat connections of $SL(N,\RR)$ theory which correspond to locally AdS metrics (in a particular gauge). We will then impose the condition of trivial holonomy to pick out those configurations which are smooth in the underlying theory.

As a starting point, consider a generalization of \eqref{b:aa} %
  \bea\label{c:ac}
A=b^{-1}\,a \,b+b^{-1}\,db~,\quad\bar A= b\,\bar a\,b^{-1}+b\,db^{-1}~,
\eea
where we express the flat gauge connections $a,\bar a$ as a linear combination of generators of weight $\pm 1$, i.e. $L_{\pm1}$ and $W^{(s)}_{\pm1}$. Note that it follows from the explicit matrix form of the generators given in (\ref{wsform}) that the generators  have non vanishing components directly above and below the diagonal. For the analysis of the holonomy it is more convenient to employ a different basis given by the $B^{(l)}_k$  matrices defined in appendix \ref{diagonalgen}
\bea\label{c:a}
 a =\left( \sum_{k=1}^{N-1} B^{(1)}_k(a_k,b_k)\right) dx^{+}~, \quad
\bar a = -\left(  \sum_{k=1}^{N-1} B^{(1)}_k(c_k,d_k)\right)  dx^{-}~,
\eea
where $B^{(1)}_k(x,y)$ is a constant $SL(N,\RR)$ matrix with entries only directly above and below the diagonal. The vielbein is given by (\ref{vielbeina}) and the metric is defined as
\bea\label{bd}
ds^2= {1\over \epsilon_N} \tr(e_{\mu}e_{\nu})  dx^\mu dx^\nu~,
\eea
with $\epsilon_N$ given in \eqref{b:bc}. We will restrict our attention to  static metrics, i.e.   $g_{++}=g_{--}$. This implies
\be\label{staticcond}
\sum_{k=1}^{N-1} a_k b_k = \sum_{k=1}^{N-1} c_k d_k ~.
\ee
Using \eqref{c:a}, \eqref{c:ac} and \eqref{staticcond}, the metric \eqref{bd} takes the following form
\bea\label{c:f}
{ds^2\over \ell^2_A}&=& d\rho^2-\left[e^{2\rho}+\Lambda_N e^{-2\rho}-2M_N\right]dt^2+\left[e^{2\rho}+\Lambda_N e^{-2\rho}+2M_N\right]d\phi^2~.
\eea
 We define
 \bea\label{c:fa}
 \beta_N& =&{1\over 2\epsilon_N}\sum_{k=1}^{N-1} b_k c_k ~,\cr
 \Lambda_N& =&  {1\over (2\epsilon_N)^2} \left(\sum_{k=1}^{N-1} b_k c_k \right) \left(  \sum_{k=1}^{N-1} a_k d_k\right)~, \cr
 M_N&=& - {1\over 2\epsilon_N} \sum_{k=1}^{N-1} a_k b_k ~,\eea
where $\beta_N$ enters into a redefinition $\rho \to \rho +\ln (\beta_N)$, which is used to bring the metric to the form \eqref{c:f}.  For any value of $\Lambda_N$ and $M_N$ the metric is asymptotically AdS. If we impose the additional requirement
\bea\label{c:gb}
\Lambda_N= M^2_N~,
\eea
the metric \eqref{c:f} is locally AdS.  This condition is equivalent to
\be\label{localcond}
 \left(\sum_{k=1}^{N-1} b_k c_k \right) \left(  \sum_{k=1}^{N-1} a_k d_k\right) = \left(   \sum_{k=1}^{N-1} a_k b_k\right)^2~.
\ee
The conditions (\ref{staticcond}) and (\ref{localcond}) are solved by demanding
\be\label{c:gc}
 b_k=\alpha\,a_k,  \quad  \quad c_k = {  \gamma} \,a_k, \quad \quad d_k ={\alpha\over  \gamma }\, a_k , \quad\quad  k=1,2,\cdots N-1~,
\ee
where $\alpha$ and $\gamma$ are real positive constants.
Further, a similarity transformation allows us to set $\alpha=\gamma=1$, and for this choice the connections (\ref{c:a}) are anti-hermitian.
Even though the solutions are all locally AdS, the higher spin fields have generically non-zero values. It is interesting that there is a gauge where all higher spin fields have vanishing stress tensor, i.e. they are truly topological matter.

In the pure gravitational theory the coefficient $M_N$ controls the ADM mass, with the exact relation being
\be
M= {M_N\over 2G}~.
\ee
And written in terms of the boundary stress tensor\footnote{We are using the conventions $L_0={1\over 2}{(H+J)}$ and $\bar L_0={1\over 2}(H-J)$;  these are the charges that generate translations on the $(t,\phi)$ cylinder. We are abusing notation, these are not to be confused with the $SL(N,\RR)$ generators.}
\be\label{gravmass}
L_0=\bar L_0={M\ell_A\over 2}={c\over 6} M_N~,
\ee
with $c$ given by \eqref{cc}. For both the gravitational $SL(2,\RR)$ connection and for $SL(N,\RR)$,  $M_N$ is proportional to the quadratic Casimir of the algebra. This motivates to extend the definition \eqref{gravmass} and define  energy for the higher spin theory as
\bea\label{c:gba}
L_0&=&{c\over 24 \epsilon_N}\tr(a^2)~, \quad
  \bar L_0 ={c\over 24\epsilon_N}\tr(\bar a^2)~.
\eea
One can check that this formula gives exactly the stress tensor of the dual CFT  for the highest weight gauge of the principal embedding of $SL(N,\RR)$, and non-principal embedding for $N=3,4$ discussed in \cite{Campoleoni:2011hg,Ammon:2011nk}.

We are now ready to analyze the spectrum of the theory using the charges \eqref{c:gba}. We first note that for solutions satisfying  \eqref{c:gc} we always have $L_0=\bar L_0$, and in terms of $M_N$ it reads
\be
L_0= \bar L_0 ={c\over 6} M_N~.
\ee
When $L_0\geq 0$ the solution corresponds to a BTZ black hole and for $L_0=-c/24$ we return again to global AdS. In the range $-c/24<L_0 <0$ there is a conical singularity at  $e^{\rho_0}=M_N$ with deficit angle $\delta= 2\pi(1-2\sqrt{-M_N})$.

In the following, we will be only interested in those configurations that do not contain horizons -- i.e. the contractible cycle is spatial and labelled by $\phi\sim \phi +2\pi$. Hence for the present ansatz \eqref{c:a} we will restrict to solutions satisfying \eqref{c:gc} and
\be\label{c:gd}
-{c\over 24}< L_0 <0~.
\ee
Solutions below this bound correspond to geometries with a conical surplus, and often  are regarded as unphysical. However,  they will play an important role in our analysis and it will be discussed in detail in the next section.

By examining the metric \eqref{c:fa}, it is tempting to declare that for the range \eqref{c:gd} the solutions are singular. As we stressed earlier, in a higher spin theory, the metric $g_{\mu\nu}$ transforms not just under diffeomorphisms but more general gauge transformations. In the Chern-Simons language the holonomy is the only meaningful observable.  Therefore we use the criterion on the holonomy   \eqref{a:b} to decide which connections \eqref{c:a}  are physically acceptable.

In our search for regular connections, we have to first demand that the matrix \eqref{c:a} is diagonalizable and has purely imaginary eigenvalues.  Note that  among all the  matrices $B^{(1)}_k$,  the ones with  $k$  odd  form a maximal commuting set. Hence  the connection which is diagonalizable can be expressed as follows
\be\label{condiag}
a_+ =\sum_{ j=1}^{\lfloor N/2\rfloor} B^{(1)} _{2j-1}(a_{2j-1},  a_{2j-1})~,
\ee
with real $a_{2j-1}$ and a similar expression for $\bar a$.  Now that the decomposition of the connection $a$ is evident, we can evaluate the holonomy \eqref{a:b}. We have
\bea\label{c:i}
{\rm Hol}_\phi (A)&=&b^{-1}\exp(\oint d\phi\, a_{\phi})\, b\cr
&\sim& \exp(2\pi \lambda_+)~,
\eea
where
\bea\label{c:j}
\lambda_{+}&=& i\diag \left ({ \pm a_{1},\pm  a_{3},\cdots    ,\pm a_{N-1} }\right)~,\quad \textrm{for even $N$}~,\cr
\lambda_{+}&=& i\diag \left ({\pm a_{1},\pm  a_{3},\cdots   , \pm a_{N-2} },0 \right)~,\quad \textrm{for odd $N$}~.\label{eigenval}
\eea
The eigenvalues for $\bar{A}$ are exactly the same as for $A$; this is another consequence of the choices in \eqref{c:gc}.
Evaluating  (\ref{c:gba}) for the connection (\ref{condiag}) gives
\bea\label{c:jaz}
L_0&=&- {c\over 12 \epsilon_N}\sum_{j=1}^{\lfloor N/2\rfloor } a_{2j-1}^2\nonumber\\
&=& {c\over 24\epsilon_N}{\tr(\lambda_+^2)}~.
\eea

From here we can now identify which solutions are smooth.
Following the discussion at the end of section \ref{classi} we remind the reader that for even $N$ both $\pm{ 1_{N\times N}}$ are regarded as the trivial element. For odd $N$ on the other hand only ${ 1_{N\times N}}$ is trivial. Hence
requiring that \eqref{c:i} exponentiates a  trivial element  is equivalent to
\be\label{trivconn}
a_{2j-1}= {n_j}, \quad\quad j=1,2,\cdots,\left\lfloor{N\over 2}\right\rfloor ~,
\ee
where $n_j$ is an integer for odd $N$.  For even $N$ there are two possibilities: either $n_j$ are all integers  which gives holonomy  $+1_{N\times N}$, or all half integers where the holonomy is $-1_{N\times N}$. Note that the deficit angle in radians $\d /2 \p$ of these solutions
is generically irrational, in contrast  to  the defects that arise in  string theory as exact orbifold backgrounds.
The condition on the mass (\ref{c:gd}) imposes constraints the $n_i$, we find
\bea\label{c:jb}
&&0<  \sum_j n_j^2  \leq {1\over 24} N(N^2-1)~.
\eea
The lower bound is set in order to discard the non-geometrical connections $A=\bar A=0$.

\medskip

To illustrate the point, lets go through a handful of values of $N$ explicitly:\\

\noindent $\bullet\;  N=2:$ Only one  half integer  or integer $n_1$ in play, and the constraint \eqref{c:jb} imposes
 \be
 (n_1)^2\leq {1\over 4} ~.
 \ee
 Hence the only half integral solution is $n_1={1\over 2}$ which corresponds to global AdS with holonomy $-1_{2\times 2}$.

 \medskip

\noindent $\bullet\;  N=3:$ There is only one integer $n_1$ and  the bound \eqref{c:jb} translates to
\be
(n_1)^2\leq 1~,
\ee
 and again their is only one solution: $n_1=1$ which is global AdS with holonomy $1_{3\times 3}$.

\medskip

\noindent $\bullet\;  N=4:$ This is a more interesting case. Here we have two  $(n_1,n_2)$, which are either integers or half integers. The mass constraint \eqref{c:jb} requires
 \be\label{d:a}
 (n_1)^2+(n_2)^2 \leq  {5\over 2}~.
 \ee
 In table {\ref{t:2} we list  smooth cases and denote the mass as well as the holonomy for each. We have three smooth connections which correspond to a conical defect. No analogous configuration exists in the purely gravitational theory with $N=2$.

\begin{table}[!h]
 \begin{center}
 \begin{tabular}{|cccc| }
 \hline
 $n_1$ & $n_2$ & $L_0$ & holonomy\\
 \hline
  ${3/ 2}$& ${1/ 2}$ &-${c/24}$ & $-1_{4\times 4}$\\
 1 & 1 &-${c/30}$  & $+1_{4\times 4}$ \\
   1 & 0 & -$c/60$& $+1_{4\times 4}$ \\
 ${1/2}$ & ${1/ 2}$ &-$c/ 120$  & $-1_{4\times 4}$\\
 \hline
 \end{tabular}
 \caption{ Set of smooth conical deficits for $N=4$ in the principal embedding. The first row is global AdS.}\label{t:2}
  \end{center}
 \end{table}

\medskip

\noindent $\bullet\;  N=5:$   Again we have two integers $n_{1,2}$ as defined in  \eqref{trivconn} which are bounded according  to \eqref{c:jb}
 \be\label{d:c}
 n_1^2+n_2^2\leq 5~.
 \ee
 In table \ref{t:1} we list all possible solutions to \eqref{d:c}; these are the smooth conical deficits of the $N=5$ theory.  \begin{table}[!h]
 \begin{center}
 \begin{tabular}{|ccc| }
 \hline
 $n_1$ & $n_2$ & $L_0$ \\
 \hline
  1& 2 &-${c/24}$ \\
   0 & 2 & -$c/30$ \\
 1 & 1&-$c/ 60$   \\
  0 & 1 & -${c/ 120}$\\
 \hline
 \end{tabular}
 \caption{ Set of smooth conical deficits for $N=5$ in the principal embedding. Again, the first row is global AdS.}\label{t:1}
 \end{center}
 \end{table}

We close with some observations for general values of $N$. Global AdS is a smooth solution that saturates the bound  \eqref{c:jb}, and the $\pm n_i$ labeling the eigenvalues are given by \eqref{b:ca}.
It is evident that, as we increase $N$, the number of allowed smooth configurations increases. To quantify this, consider the bound \eqref{c:jb} for odd $N$
\be
 \sum_{i=1}^{\lfloor N /2 \rfloor} n_i^2\leq {1\over 2} \epsilon_N= {1\over 24} N(N^2-1)~,\label{treshold}
\ee
and there is an analogous bound for even $N$. The problem is reduced to counting lattice point in $N$ dimensional space inside a ball of radius $\sqrt{\epsilon_N}$. For large $N$, $\epsilon_N \sim N^3$ and the number of solutions scales as $N^{3N/2}$.
In the limit $N\to \infty$ the set of allowed smooth $L_0$ becomes dense in the interval $[-{c/24},0]$.

\subsection{The conical deficit as a  wormhole}

As discussed in section \ref{sec:hol}, the metric is not invariant under higher spin gauge transformations which are given by Chern-Simons gauge transformations involving the generators $W^{(s)}_m$.  In the following we show that the connection constructed in the previous section  can be  gauge transformed such that the conical defect metric becomes a smooth wormhole metric.  This is analogous to the situation of black hole  solutions  carrying higher spin charge discussed in
\cite{Gutperle:2011kf,Ammon:2011nk}, where a gauge exists in which the black hole metric becomes  a smooth traversable wormhole.

For definiteness we choose the case of odd $N$.
The gauge transformation to a wormhole gauge can be constructed using constant block matrices with  parameter $\gamma_k$.
\be
[\Lambda_{k}(\gamma_k)]_{ij}=  \delta_{i,{2k-1}} \delta_{j,{2k-1}}+ \delta_{i,{2k}} \delta_{j,{2k}}+ \gamma_k  \delta_{i, 2k-1} \delta_{j,2k}~.
\ee
The gauge transformation we will employ is given by
\be
\Lambda= \sum_{k}^{\lfloor N/2\rfloor} \Lambda_{k}(\gamma_k)~.
\ee
The new gauge connection is given by
\bea
A' &=&  \Lambda^{-1}  \big( b^{-1} a\, b   + b^{-1} d\, b\big) \, \Lambda~, \nonumber \\
\bar A'&=&   \Lambda \, \big( b^{}\, \bar a \,b^{-1}  + b^{}\, d\, b^{-1}\big) \, \Lambda^{-1}~,
\eea
where $a$ is given by (\ref{condiag}). Note that the vielbein component $e_\rho$ remains unchanged after the gauge transformation.
To avoid cross terms in the metric of the form $g_{\rho \pm}$ one demands that
$\tr(L_0 A_+')= \tr(L_0 \bar A_-')=0$, which implies
\be
\sum_i \gamma_i n_i =0~.
\ee
The  metric components $g_{++}$ and $ g_{--}$ are also unchanged and given by
 \be
 g_{++} =g_{--} = -{1\over 2 \epsilon_N} \sum n_i^2~. \\
  \ee
 The $g_{+-}$ component is affected by the gauge transformation and it reads
 \be
 g_{+-}= -{1\over 4 \epsilon_N} \tr(A'_+\bar A'_-)= -{1\over 4\epsilon_N}\sum_i   \Big\{ n_i^2 \big(e^{-2\rho} +  e^{2\rho}\big)  +4 \gamma_i^2 n_i^2  \Big\}~.
\ee
Note that we have not shifted the $\rho$ coordinate in order to make the argument simpler. The $\phi$ and $t$ components of the metric become

\be
g_{tt}= - {1\over 2 \epsilon_N}  \sum_i\Big\{  n_i^2  (e^\rho + e^{-\rho})^2+ 4 \gamma_i^2 n_i^2\Big\}
\ee
and
\be
g_{\phi\phi} = {1\over 2 \epsilon_N}  \sum_i \Big\{ n_i^2 (e^\rho - e^{-\rho})^2 + 4 \gamma_i^2 n_i^2\Big\}~.
\ee
The component $g_{\phi\phi}$ is the sum of two squares and it does cannot vanish unless all the $\gamma_k$ are zero. Hence after the gauge transformation the $\phi$ circle never closes off and the metric is a traversable wormhole.

\subsection{Other embeddings}

The smooth connections in section \ref{smoothdef}   were constructed  using the principal embedding, as evidenced by the normalization of the metric (\ref{bd}) and the form of the smooth global AdS connection in (\ref{b:aa}).  Consequently we can view the conical defects as excitations above the global AdS vacuum, where the conformal dimension of the dual states in the CFT is given by  (\ref{c:gba}).
One might ask wether this construction can be generalized to accommodate a non-principal embedding, or if their are other choices of vacuum.

The first generalization  is to keep the normalization and the radial gauge the same as before, but  to consider connections with non zero entries further away from the diagonal.
We consider the following ansatz
\bea\label{c:anew}
 a &=&\left( \sum_{k=1}^{N-m} B^{(m)}_k(a_k,b_k)\right) dx^{+}, \quad
\bar a = -\left(  \sum_{k=1}^{N-m} B^{(m)}_k(a_k,b_k)\right)  dx^{-}~,
\eea
where  $m>1$ and $a_k,b_k$ are constant parameters.
 Using the properties of $B^{(l)}_k$ given in appendix \ref{diagonalgen}, it is straightforward to show that the metric, after a shift in the radial coordinate, is given by
\bea\label{e:f}
{ds^2\over \ell_{A}^2}= d\rho^2-\left(e^{m\rho}+ M^{(m)}_{N}e^{-m\rho}\right)^2 dt^2+\left(e^{m\rho}-M^{(m)}_{N} e^{-m\rho}\right)^2 d\phi^2~,
\eea
 with
 \bea\label{c:bnew}
 M^{(m)}_N= -{1\over 2\epsilon_{N}}  \sum_k a_k b_k~.
  \eea
The metric is a locally AdS, and  the curvature radius of  (\ref{e:f}) is rescaled by a factor  $1/m^2$ compared to the  scale in \eqref{b:a}.  Furthermore, for particular values of the parameters (\ref{c:anew}) we can produce  global AdS by setting \eqref{c:bnew} to minus a quarter.  But in general the holonomy of this larger version of global AdS  is not trivial, hence it would correspond to a false vacuum. Most importantly, a smooth connection in this gauge  should not be interpreted as an excitation around this false vacuum. Since any smooth $SL(N,\RR)$ connection can be brought to the form \eqref{c:a}, the appropriate interpretation is as an excitation of the global AdS associated to the principal embedding.

The second generalization is to consider a non principal embedding labelled by a partition $P$ of  $N$. The notation and conventions we use for the non principal embeddings can be found in appendix \ref{nonprinc}.
For a given embedding $P$ the  metric  is normalized as follows
\bea
ds^2 = {1 \over \e_P} \tr( e_\mu^P e_\nu^P) dx^\mu dx^\nu~,
\eea
where there is a suitable definition of the vielbein  $e_\mu^{P}$ as a function of $(A,\bar A)$, and $\e_P$ is given by (\ref{epspdef}).\footnote{The construction of metric like fields changes depending on the matter content of the embedding. See e.g. \cite{Castro:2011fm} for a recent discussion.}
We consider the analogous ansatz to (\ref{c:ac}) for the non-principal embedding
\be
A=b_P^{-1}\,a_+\,b_P \;dx^+ +b_P^{-1}\,db_P, \quad \bar A=b_P^{}\,\bar a_+\,b_P^{-1} \;dx^- +b_P^{}\,db_P^{-1}, \quad b_P = \exp(\r L_0^{(P)})~.
\ee
Global AdS is represented by
\be
a_+ = L_1^{(P)} +{1\over 4}L_{-1}^{(P)}, \quad \bar a_- = -(L_{-1}^{(P)} +{1\over 4}L_{-1}^{(P)})~,
\ee
Where the $SL(2,\RR)$ generators  $L^{(P)}_{0,\pm1}$ are defined in (\ref{slbloack}).
Upon Drinfeld-Sokolov reduction the non principal embeddings realize extended conformal symmetry algebras other than the $\W_N$ algebra. For example  the $P=2+1$ embedding of the $N=3$ theory realizes the $\W_3^{(2)}$  algebra \cite{Bershadsky:1990bg,Polyakov:1989dm}.
A detailed discussion of the  embeddings and their associated chiral algebras can be found in \cite{Bais:1990bs}.

In contrast to the principal embedding case, the global AdS connection does not exponentiate to the trivial element. Strictly speaking, it is trivial only when the representations appearing in the branching
are all even-dimensional (for N even) or all odd-dimensional (for N odd), i.e. when $P_{2A+1}=0$ for $N$ even or $P_{2A}=0$ for $N$ odd. In terms of the associated field content in the bulk  the  global AdS connection is smooth if and only of there are no half-integer spin fields.

One can construct smooth conical defects utilizing the the matrices  $B^{(1,A)}_k(x,y)$ for each $A\times A$ block appearing in the partition $P$, generalizing the analysis of section  \ref{smoothdef} in a straightforward manner. We leave details   as an exercise to the reader.

\section{Spectrum of surpluses and other smooth bulk configurations}\label{sec:CSur}

In the previous section we described the conditions in the $SL(N,\RR)$ Chern-Simons formulation  required for a solution  to be smooth and to have a locally AdS metric.
In this section, we will systematically construct the spectrum of the theory and determine the higher spin charges carried by each smooth configuration. In doing so, we will be led to  generalize the discussion of the previous section in two ways.

First of all, in (\ref{c:gd}), we explicitly restricted ourselves to solutions whose energy lies above the global AdS value of $-{c\over 24}$, i.e. $M_N \geq -{1\over 4}$. If we relax this condition, we can still find smooth holonomy configurations labelled by integers $n_i$, but no longer obeying the constraint (\ref{c:jb}). In fact, the solutions with $M_N < -{1\over 4}$ correspond to $n_i$ satisfying
\bea\label{surplint}
&& \sum_j n_j^2 > {1\over 24} N(N^2-1)~.
\eea
From the expression $\delta= 2\pi(1-2\sqrt{-M_N})$, we see that these solutions correspond to conical surpluses (i.e. with negative $\delta$). Being below the global AdS solution,
these solutions are probably unphysical from the point of view of the
$SL(N,\RR)\times SL(N,\RR)$ theory.
Nevertheless, we investigate under what conditions they
obey the standard falloff conditions on  the Chern-Simons connection that are appropriate for the Drinfeld-Sokolov reduction.

A second generalization is to consider smooth configurations in the Euclidean version of higher spin gravity, which is described
by an $SL(N,\CC )$ Chern-Simons theory. We will see that, due to the
absence of a certain reality condition and the fact that $SL(N,\CC )$  has a $\ZZ_N$ center, the spectrum of admissible configurations is now considerably larger than in the Lorentzian theory.

\subsection{Falloff conditions and conical surpluses}\label{sec:4bulk}

First let's recapitulate some of the properties of our smooth solutions. As explained in section \ref{sec:hol} and \ref{sec:CD},
after imposing suitable boundary -- and gauge-fixing conditions -- a general flat Chern-Simons connection can be written as
\bea
A=b^{-1}\,a\,b+b^{-1}\,db~,\quad\bar A= b\,\bar{a}\,b^{-1}+b\,db^{-1}~,\label{Aitoa}
\eea
where $b = e^{\r L_0}$, $a=a_+ dx^+$ and $\bar a = \bar a_- dx^-$  are the  $SL(N,\RR)$ Lie algebra elements \eqref{condiag}. All of these solutions correspond to locally AdS$_3$ metrics, which automatically obey the standard Brown-Henneaux asymptotic falloff conditions \cite{Brown:1986nw}.
However, in the Chern-Simons formulation of the theory, the falloff condition is to be imposed on the connection rather than the metric.
The standard falloff condition on the Chern-Simons connection is the one proposed in \cite{Campoleoni:2010zq}
\be
\left( A - A_{\rm AdS} \right)_{|\rho \to \infty}  = \bigO (1)~,\label{asAdS}
\ee
with $A_{\rm AdS}$ the connection \eqref{b:aa} for global AdS. There is a similar condition for $\bar A$ as well.
The above constraint is the starting point for identifying the asymptotic symmetries of the theory. It imposes
a Drinfeld-Sokolov (DS) condition on the connection, and leads to the identification of the asymptotic symmetry algebra as a  ${\cal W}_N$-algebra.

One can easily see that our smooth solutions in the block-diagonal gauge (\ref{condiag}) don't lie on the DS constraint surface (\ref{asAdS}). We will now investigate under what conditions they
can be brought into the form (\ref{asAdS}) by an $SL(N,\RR )$ gauge transformation.  We will not construct the explicit gauge transformation. Instead we will re-visit the smoothness condition imposed by the holonomy in the DS gauge and compare with the results in section \ref{smoothdef}. This will give us a criterion for the block diagonal solutions to obey the DS boundary conditions. We will see that this criterion is satisfied only for a special class of conical surpluses, and not the deficits.

Using (\ref{b:aa}), the asymptotic condition (\ref{asAdS}) is equivalent to imposing
\be
a_+ = L_1 + u~,
\ee
where $u$ is an upper triangular matrix and similar for $\bar a_-$.
As explained in \cite{Balog:1990mu,Campoleoni:2010zq}, residual gauge transformations allow us to further fix the form of $u$. A particularly useful gauge for computations is the one
where $u$ has nonzero entries only in the first row. Making a further change of basis, we can assume $a_+$ to be of the form
\be
a_+=
\left(
\begin{array}{cccccc}
 0  & u_1 & u_2 & \cdots & 0 & u_{N-1}    \\
-1  & 0  & 0  &\cdots &0 & 0  \\
 & & \cdots&&\\
 0  & 0  &0  &\cdots & -1 & 0
\end{array}
\right)~.\label{quadrbasis}
\ee
In this gauge, the characteristic polynomial of $a_+$ has the simple form $\l^N + u_1 \l^{N-2} - u_2 \l^{N-3} +\ldots + (-1)^{N} u_{N-1}$.
It is a well known property that, as long as the eigenvalues are nondegenerate, a matrix of the form (\ref{quadrbasis}) is diagonalized by a Vandermonde matrix $V$:
\be\label{vander}
V=
\left(
\begin{array}{cccc}
\l_1^{N-1}  & \l_2^{N-1}  & \cdots & \l_{N}^{N-1}  \\
& & \cdots&\\
\l_1  & \l_2  & \cdots & \l_{N}  \\
  1  & 1  & \cdots & 1
\end{array}
\right)~,
\ee
where the $\l_i$ are the eigenvalues. This decomposition completely characterizes all solutions that obey the boundary condition \eqref{asAdS}. From here it is evident that when the eigenvalues are degenerate, $a_+$  and $\bar a_-$ in the DS gauge are not diagonalizeable.

Following the analysis of the previous section, the smooth solutions in the DS gauge will again be are characterized by $N$ (half) integers $n_j$ related to the eigenvalues $\l_j$ of $a_+$ and $\bar a_-$
\be i\, n_j = \l_j ~.\label{lambdavsn}\ee
We will  assume in this section that the $n_i$ form an ordered sequence $n_1 \geq n_2\geq \dots \geq n_N$.
As we derived in \eqref{trivconn}, for $N$ odd we have $n_i\in \ZZ$ while for $N$ even either $n_i\in \ZZ$ or $n_i\in \ZZ/2 \backslash \{ 0 \}$. Because $a$ and $\bar a$ are real connections, they are subject to the additional constraint
\be
n_i = - n_{N+1 - i}~.\label{constrn}
\ee
And in addition, demanding that the matrices are diagonalizable requires $n_i\neq n_j$ for all $i\neq j$.

To ease the notation, we define the vector $n=(n_1,\cdots, n_N)$ and we will denote by $\rho=(\r_1,\cdots, \rho_N)$ the values corresponding to the global AdS solution in \eqref{b:da}
\be
\r_i \equiv   {N +1\over 2} - i~.\label{Weyl} \ee
Note that $\r $ is the Weyl vector $SU(N)$ and that $\r^2 = \e_N $.  In this notation, the general solution has energy
\be\label{L0sec4}
L_0 = \bar L_0 = -{c \over 24 \e_N}  n^2~,
\ee
according to our definition in \eqref{gravmass}. Configurations with $n^2 <\r^2$ correspond to conical deficits metrics, which are the solutions we discussed in section \ref{smoothdef}; while those with $n^2 >\r^2$ describe conical surpluses. The conical surpluses seem to be unphysical configurations  in the description of the $SL(N,\RR)\times SL(N,\RR)$ higher spin theories because their energy is unbounded from below. However, we will keep an open mind and not throw them away.

For those conical defects with $n^2 < \r^2$ it's easy to see that some of the eigenvalues must be degenerate. Hence any connection that satisfies  (\ref{asAdS}) and has the same eigenvalues as the conical defect connections constructed in section \ref{smoothdef} is non-diagonalizeable. It is therefore not possible to gauge-transform   the conical defect  from the diagonalizeable block-diagonal form to (\ref{asAdS}).

The smooth connections for which none of the eigenvalues coincide can be brought to the form (\ref{asAdS}) and the block diagonal form \eqref{condiag}. These consist, in addition to global AdS, of those conical surpluses where $n^2>\rho^2$ and $n_i\neq n_j$.

Even though all of these connections will exponentiate to the trivial element, there is one aspect that makes them distinct from the vacuum AdS solution:  the conical surpluses can carry non-trivial higher spin charge. Using the highest weight gauge,
where only the highest weight generators $W_{-(s-1)}^{(s)}$ are turned on, we have
\be
a_+ = L_1 + {1\over k_{\rm cs}} \sum_{s=2}^{N} {k_{\rm cs}^{-{s-2\over 2}}\over t^{(s)}_{s-1}}  w^{(s)}_0 W_{-(s-1)}^{(s)}~,
\label{hwgauge}
\ee
where $w^{(s)}_0$ is the spin $s$  charge,   and in particular the eigenvalue \eqref{L0sec4} is given by $L_0=w^{(2)}_0$. The normalization $ t^{(s)}_{s-1} \equiv \tr( W_{(s-1)}^{(s)} W_{-(s-1)}^{(s)})$ is given by  (\ref{app:tr}) and $k_{\rm cs}$ is the Chern-Simons coupling (\ref{cc}).
We have chosen a specific normalization of the higher spin charges for later convenience. For example, for $N=3$ our normalization leads to the
 Poisson brackets of the classical $\W_3$ algebra normalized as  in (\ref{PoissW3}).

To determine the coefficients $w^{(s)}_0$, it suffices to demand that $a_+$ has the correct eigenvalues (\ref{lambdavsn}).
This will be the case if and only if the $N-1$ independent trace invariants take the values
\be
{(- i)^s \over s} \tr (a_+)^s = {1 \over s} \sum_{i=1}^N (n_i)^s \equiv C_s (n)~, \qquad s = 2,\ldots, N~.
\ee
Plugging in (\ref{hwgauge}), we get
\be
{1 \over s} \tr \left(  L_1 + \sum_{s=2}^{N} {k_{\rm cs}^{-s/2}\over t^{(s)}_{s-1}}  w^{(s)}_0 W_{-(s-1)}^{(s)} \right)^s = i^s C_s (n)~.
\ee
 The trace picks out the weight zero terms in the expansion. Working this out by using \eqref{wsform}, together with  \eqref{sl2comms} and \eqref{app:tr}, leads to a unique solution for  the $w^{(s)}_0$ which can be found recursively. Up to spin four we have
 \bea\label{surplcharge}
 w^{(2)}_0  &=&- k_{\rm cs} C_2(n)~,\nonumber\\
 w^{(3)}_0 &=& - i k_{\rm cs}^{3/2} C_3 (n) ~,\nonumber\\
 w^{(4)}_0 &=&
  k_{\rm cs}^2 \left( C_4 (n) - {C_4(\r) \over C_2(\r)^2 } C_2(n)^2\right)~.\label{charges}
 \eea
One quick check is that global AdS ($n_i = \r_i$) has vanishing higher spin  $(s>2)$ charges.
For general spins, one can also easily work out the coefficient of the  term proportional to $C_s (n)$:
\be
  w^{(s)}_0 = i^s k_{\rm cs} ^{s/2}  C_s (n) + \ldots ~.\label{largeNcharges}
\ee
Since the $n_i$ satisfy (\ref{constrn}) the $C_s(n)$ vanish for odd $s$. As a consequence, all charges $w^{(s)}_0$ with $s$ odd vanish for our configurations, as required by the reality of $a_+$.

\subsection{Admissible connections in $SL(N,\CC)$}
As will become clear in section \ref{surplvsyoung}, the conical surpluses will only account for a subset  of the light primaries of the two dimensional CFT. Among other things, this is an artifact of reality constraints of  the $SL(N,\RR)\times SL(N,\RR)$ connections.  A more complete analysis  requires generalizing the above construction of admissible solutions to the Euclidean signature higher spin theory, in which the gauge group is $SL(N, \CC)$. Since the analysis for the Euclidean theory is analogous to the above discussion, we will only highlight the relevant differences.

In the Euclidean theory, the connection $A$ takes values in the Lie algebra of $SL(N,\CC )$. Following the conventions of  \cite{Banados:1998gg}, we take $\bar A$ to be anti-Hermitean conjugate of $A$, i.e.
$ \bar A = - A^\dagger$.\footnote{This condition follows naturally from a Wick rotation of the time coordinate. For example, the connections in (\ref{b:aa}) satisfy this property
 after analytic continuation $t \to t= i t_E$ and using (\ref{herm}).} In analogy with (\ref{Aitoa}) we will consider flat connections of the form
\bea
A&=b^{-1}\,a_+ \,b\, dz +b^{-1}\,db~,\qquad\bar A&= -  b\, (a_+)^\dagger \,b^{-1}\, d \bar z +b\,db^{-1}~,
\eea
where $z=\phi +it_E$, $b = e^{\r L_0}$  and $a_+$ is now an $SL(N,\CC )$ Lie algebra element.

For $SL(N, \CC)$ Chern-Simons theory,  instead of (\ref{vielbeina}), the vielbein and spin connection are given by
\be\label{vielbeincomp}
e={\ell_{A}\over 2 i} \big(A-\bar A\big)~, \quad \omega= {1\over 2}\big( A+\bar A\big)~,
\ee
and the formula (\ref{metric}) for the metric is to be replaced  by
\be
g_{\mu\nu} = - {1\over \epsilon_N}  \tr\big (e_\mu e_\nu\big)~.
\ee

The holonomy matrix \eqref{a:b} is now an $SL(N, \CC)$ matrix. Demanding that the holonomy is trivial means that it lies in the centre of  group. For $SL(N, \CC)$ the center is given by
\be
e^{-{2\pi i {m\over  N}}} \,{ 1}_{N\times N} ~, \quad m\in \ZZ_N~.
\ee
Thus, smooth euclidean solutions are labeled by eigenvalues $\lambda_i$ which read 
\be
\l_j =  i (m_j - {m\over N}) \equiv  i\; n_j^{\prime} ~,
\label{lambdaslnc}
\ee
with $m_j$ an integer. This should be compared with Eqs. (\ref{c:i}, \ref{c:j}, \ref{trivconn}).
 Therefore, in contrast to the admissible solutions for $SL(N,\RR)$ -- labelled by (\ref{lambdavsn}) -- now a $N$th root of unity is allowed.

In addition, a connection  is no longer subject to the reality constraint (\ref{constrn}), which was an earlier  consequence of the connections $(A,\bar A)$ being real. However, they are still subject to the $SL(N)$ constraint on the determinant of the holonomy matrix. This implies that the sum $\lambda_i$ vanishes, and hence
\be
m=\sum_i m_i~.
\ee

When the eigenvalues \eqref{lambdaslnc} are nondegenerate, the connection can be brought to a gauge where it satisfies \eqref{asAdS} and therefore is of the form \eqref{hwgauge}. The remaining computation of the higher spin charges goes through in the same way with the replacement of $n_i$ by $n_i^{\prime}$. Notice that because of the absence of the constraint (\ref{constrn}), the odd spin charges $w_0^{(s)}$ no longer vanish for a generic smooth configuration.

Summarizing, for the Euclidean theory we have additional smooth solutions in the spectrum coming from the non-trivial center of $SL(N,\CC)$ and the absence of the reality condition  \eqref{constrn}. The spacetime interpretation of these solutions is also modified.  Because the eigenvalues are not necessarily paired up, in general they cannot be brought to a block diagonal form as in \eqref{c:a}. This implies that those $SL(N,\CC)$ solutions that do not satisfy   \eqref{constrn} are not conical defects.

\section{Relation to the light primaries in the ${\cal W}_N$ CFT}\label{sec:WN}

The smooth configurations in the previous section seem to be configurations one would not want to include in the description of the $SL(N)$ higher spin theories because their energy is unbounded from below. However, we will propose an intriguing role for them in the analytic continuation of the dual to the ${\cal W}_N$ minimal model CFTs. We will see that the smooth solutions are related by this analytic continuation to an interesting class of primaries in the ${\cal W}_N$ minimal models -- the so-called light primaries -- which become arbitrarily light in the large $N$ limit. We find that the discrete spectrum of these primaries matches with those of the surpluses, and their Euclidean generalizations,
even at finite $N$.
We will also check that the spin 3 as well as spin 4 charges also agree in a very nontrivial way.

The duality of \cite{Gaberdiel:2010pz} relates the 't Hooft limit of the $k$-th ${\cal W}_N$ minimal model to the Vasiliev higher spin theory $hs[\lambda]$ with two complex scalars.  An important aspect of this proposal is the implementation of the 't Hooft limit. Recall that the limit is given by   $N,k\to \infty$ while the coupling
\be
\lambda= {N\over k+N} \leq 1~,
\ee
is fixed.  This limiting procedure was crucial for subsequent checks of the duality, and it affects the finite $N$ realization of it. This is an essential point which we will now review.

The global (or rigid) symmetry of the $\W_N$ minimal model, i.e. the wedge algebra that generates $\W_N$, is $SL(N,\RR)$ in Lorentzian signature ($SL(N, \CC)$ in Euclidean signature). In the duality, this global symmetry is mapped to the gauge group of the bulk.

The analysis in
\cite{Gaberdiel:2011wb, Gaberdiel:2011zw} showed that in the 't Hooft limit the spectrum of the $\W_N$ minimal model instead falls into representations of $\W_\infty[\lambda]$, and hence the gauge group in the bulk is its wedge algebra, $hs[\lambda]$. This is seemingly different from the large $N$ limit of $SL(N)$. However, a key observation in  \cite{Gaberdiel:2011zw}  was to interpret this in terms of the level-rank duality of the coset description of the CFT in the 't Hooft limit. More precisely, it is believed that
\be\label{clr}
{SU(N)_k \otimes SU(N)_1 \over SU(N)_{k+1}}\equiv {SU(M)_{k'} \otimes SU(M)_1 \over SU(M)_{k'+1}}~,
\ee
and the dual rank and level is given by
\be
M={N\over N+k}~,\quad k'={M\over N}-M~,
\ee
which are not integers in general.
On the right-hand side of \eqref{clr},  taking the 't Hooft limit corresponds to an analytic continuation of  $M\to \lambda$ with $0\leq M\leq1$. The central charge is the same on both sides of \eqref{clr}. This is an indication that the relevant global symmetry of the right-hand side is the extension of  $SL(M)$ to non-integer value $\lambda$, i.e. $hs[\lambda]$. It was also shown that the spectrum of a large class of primaries on both sides of (\ref{clr}) matched  under the continuation $N\to -\lambda$.

 We now review the analysis of the spectrum of the CFT and its relation to the bulk excitations. The primaries in the  ${\cal W}_N$ CFT are labelled by two representations $(\Lambda_{+}, \Lambda_{-})$ of $SU(N)$. It is easy to identify the states $(0,\Lambda)$ and $(\Lambda, 0)$ with various perturbative multi-particle excitations of the scalars in the bulk theory. However, the states $(\Lambda,\Lambda)$ are more puzzling. For small representations $\Lambda$ these primaries have a dimension $\sim {1\over N}$ (see below) and are therefore very light states in the bulk. It was argued in
\cite{Gaberdiel:2011zw} that these light states become null in the strict $N=\infty$ limit and decouple from the spectrum. However, at large but finite $N$, this is not so and evidence was presented \cite{Papadodimas:2011pf} from computation of the four point function that these states do appear even in tree level diagrams to leading non-vanishing order in ${1\over N}$.
More generally, the (not necessarily light) states of the form $(\Lambda,\Lambda)$ form a discretuum in the large $N$ limit which has exponential degeneracy, though the majority of them decouple from perturbative correlators due to fusion rules of the CFT.

Thus it is important to understand what these puzzling light states are from the bulk point of view. The discretuum of conical defects we have found in the $SL(N)$ theories (for large $N$) seems to have a similar flavor. Below we will argue that a generalization of the analytic continuation  of \cite{Gaberdiel:2011zw} relates the puzzling "light primaries" to conical surpluses (and their generalizations in the Euclidean $SL(N, \CC)$ theory). The spectra as well as the spin 3 and 4 charges, of the two sides match precisely even at finite $N$! While the analytic continuation between the two sides remains somewhat mysterious,
we take the non-trivial agreement between the spectra as strong indication that it is on the right track.

\subsection{The spectrum of light states}
Recall that the spectrum of primaries labeled $(\Lambda,\Lambda)$ in the $k$-th $\W_N$ minimal model CFT is given by operators with scaling dimension
\be
h(\Lambda, \Lambda)= {\C_2(\Lambda)\over (N+k)(N+k+1)}~.
\ee
See e.g.  (2.10) and (3.6) of \cite{Gaberdiel:2010pz}. This is an exact expression -- no large N limit has been taken.
Here $\C_2(\Lambda)$ is the quadratic Casimir of the $SU(N)$ representation $\Lambda$.
Let us write this in terms of the central charge
\be
c_N(k)=(N-1)[1-{N(N+1)\over (N+k)(N+k+1)}]=(N-1){k(2N+1+k)\over (N+k)(N+k+1)}~.
\ee
We can then write
\be\label{light}
h(\Lambda, \Lambda)= c_N(k){\C_2(\Lambda)\over (N-1)k(2N+1+k)}~.
\ee

The generalized level rank duality mentioned above was studied in the 't Hooft limit
in \cite{Gaberdiel:2011zw}, and involved an analytic continuation of $N$ to $-\lambda$.
Here we will take a slightly different point of view. We will study the ${\cal W}_N$ theories at {\it finite} $N$ and consider the analytic continuation of the level $k$ from positive integers to $k=-(N+1)$. Note from the definition of $\lambda$ that this corresponds to
taking $\lambda=-N$. We propose that this continuation maps the bulk dual of the
${\cal W}_N$ minimal model to the one described by $SL(N)$  higher spin theory.

Let us carry out the continuation $k=-(N+1)$ on the spectrum (\ref{light}), leaving the central charge $c_N(k)=c$ as it is.
Then
\be\label{hLL}
h(\Lambda ,\Lambda)= -c{\C_2(\Lambda)\over (N-1)(N+1)N} =-{c\over 12 \epsilon_N}\C_2(\Lambda)~,
\ee
with $\epsilon_N$ given by \eqref{b:bc}.

In order to compare with the bulk states in section \ref{sec:4bulk}, we need to arrange \eqref{hLL} in a more suitable form.  Recall that we can write  the quadratic Casimir of $SU(N)$ as $ \C_2(\Lambda)={1\over 2}(\L, \L + 2 \r )$.
In terms of Young diagram data,  the  components of  $\L$ are
\be
\L_i =  r_i - {B\over N}~,
\ee
where $r_i$ is the number of boxes in the $i$-th row and $B = \sum_i r_i$ is the total number of boxes.
We can then rewrite the  quadratic Casimir as
\be
\C_2(\Lambda)= {1\over 2}(\sum_i \tilde n_i^2-\epsilon_N)~,
\ee
 where the $\tilde{n}_i$ are distinct numbers given by
\be
\tilde n_i = \L_i + \r_i = r_i + {N+1 \over 2} - i - {B \over  N}~.\label{ntildes}
\ee
Adopting cylinder normalization, the weight \eqref{hLL} is related to the eigenvalue of $L_0$ by the shift $w_0^{(2)}=h-{c\over 24}$. After taking this into account,
we see that the conformal weights \eqref{hLL} and \eqref{surplcharge}  match if we identify $\tilde n_i = n_i$. Or comparing with the Euclidean signature surpluses, whose eigenvalues are given in (\ref{lambdaslnc}), we would identify
${\tilde n_i} = n_i^{\prime}$. Before looking more closely at these identifications, let us  compute the higher spin charges of the light primaries.

\subsection{Higher spin charges}

On the CFT side, the higher spin charges of the $(\L, \L)$ primaries can be computed (at least for low values of the spin) following the method of appendix
C in \cite{Gaberdiel:2011zw}. A closed form expression for the charges is known in a particular non-primary basis, the Miura basis, which is
related to the primary basis by a nonlinear field redefinition. The non-primary zero mode eigenvalues are
\be
u^{(s)}_0  = (-1)^{s-1} {\alpha_0}^s \sum_{i_1< \ldots <i_s} \prod_{j=1}^s \left( \L_{i_j} + ( s - j) \right)~,
\ee
where
\be \alpha_0 = {1 \over \sqrt{ (k+N) (k+N+1)}} ~.\ee
As in the previous subsection, it will be useful to work out the zero modes in terms of the shifted vector $\tilde n = \L + \r$ given in
(\ref{ntildes}).

Up to spin four, one finds after considerable algebra
\bea
u^{(2)}_0  &=& \a_0^2 \left[C_2 (\tilde n) - C_2(\r)\right] ~,\cr
u^{(3)}_0  &=& \a_0^3 \left[ C_3(\tilde n) - (N-2) \left(C_2 (\tilde n) - C_2(\r )\right)\right]~,\cr
u^{(4)}_0  &=& \a_0^4 \Big[ C_4(\tilde n) - \half C_2(\tilde n)^2 - {3 \over 2} C_3(\tilde n) ~,\cr
&& + {(N-3)(N-2)(N+23) \over 24}
C_2(\tilde n) -{1 \over 48} (5 N + 223) \left( \begin{array}{c} N+1 \\ 5 \end{array} \right)\Big]~.\label{ucharges}
\eea
The transformation to  the charges in the primary basis is given  by \cite{DiFrancesco:1990qr}\footnote{As in e.g. \cite{Bilal:1994js}, we use conventions where the
classical Poisson brackets of the modes are as in (\ref{classPoiss1}, \ref{classPoiss2}), and the quantum commutators are obtained by replacing $i \{\ ,\ \}_{PB } \to [\ ,\ ]$ as in (\ref{qcomm1}, \ref{qcomm2}). To compare
with \cite{DiFrancesco:1990qr} one has to replace $w^{(s)\, here} \to - w^{there}_s$ and $u^{(s) here}\to - a^{there}_s$. In particular, this
changes the sign of the nonlinear term in $w^{(4)}_0$. }
\bea
w^{(2)}_0 &=& u^{(2)}_0-{c\over 24}~, \cr
w^{(3)}_0 &=& u^{(3)}_0  + (N-2) \a_0 u^{(2)}_0~, \cr
w^{(4)}_0 &=& u^{(4)}_0  + {3\over 2} (N-3) \a_0 u^{(3)}_0  + { 3 \over 5}(N-3) (N-2)\a_0^2 u^{(2)}_0 \cr && + {(N-3)(N-2)(5 N + 7)\over 10 N (N^2 -1)} (u^{(2)}_0)^2~.
\eea
Plugging in (\ref{ucharges}) one gets many cancelations and the end result is
\bea
w^{(2)}_0 &=& \a_0^2 C_2 (\tilde n) ~, \nonumber \\
w^{(3)}_0&=& \a_0^3 \;C_3(\tilde n)~,\nonumber \\
w^{(4)}_0 &=& \a_0^4 \left( C_4 (\tilde n) - {C_4(\r) \over C_2(\r)^2 } C_2(\tilde n)^2\right)~.\label{chargesCFT}
\eea
The calculation of the $s>4$ higher spin charges quickly gets quite cumbersome. However, one can easily extract the coefficient of the term proportional
to $C_s (\tilde n)$:
\bea
w^{(s)}_0 &=&  u^{(s)}_0 + \ldots = (-1)^{s-1} \a_0^s \sum_{i_1< \ldots <i_s} \tilde n_{i_1} \tilde n_{i_2} \ldots \tilde n_{i_s} + \ldots\nonumber \\
&=&  \a_0^s  C_s (\tilde n) + \ldots ~.\label{largeNchargesCFT}
\eea

Now we can compare (\ref{chargesCFT}), (\ref{largeNchargesCFT}) with  the charges on the gravity side (\ref{charges}), (\ref{largeNcharges}).
First we observe that, for $N=3$, the normalizations of the higher spin generators implicit in (\ref{ucharges},\ref{chargesCFT})
matches with the normalization (\ref{hwgauge}) chosen on the gravity side. We show this explicitly in appendix \ref{appnorms} by comparing the $\W_3$ commutation relations on both sides.
As discussed in the previous subsection, to compare the CFT calculation with the gravity side  we have to replace in the above expressions
\bea
\a_0 = \sqrt{{c\over (N-1)k(2N+1+k)}}\,\to\, i \sqrt{ c \over 12 \e_N} = i \sqrt{k_{\rm cs}}~.\label{CAC}
\eea
Doing this, one sees that all the charges computed so far match precisely upon the identification of $\tilde n_i$ with $n_i$ (or with $n_i'$ in the Euclidean signature).

It is quite nontrivial that  both sides of the computation match without taking any large $N$ limit. This seems to suggest a
way of obtaining higher spin charges in the ${\cal W}_N$ models from the much simpler gravity  computation.
We also can take the agreement as an indication that on the bulk side as well, we can infer results about the $hs[\lambda]$ higher
spin theory from the $SL(N,\RR)$ theory via the analytic continuation of the former by setting $\lambda=-N$.

\subsection{Matching the primaries with smooth connections}\label{surplvsyoung}
We showed above that, up to spin four at least, the charges of bulk smooth configurations, characterized by $n_i$ (or $n_i^{\prime}$), match with
those of $(\L,\L)$ primaries in the  ${\cal W}_N$ minimal models, characterized by $\tilde n_i$.
This required the identification of $\tilde n_i$ with $n_i$ (or $n_i^{\prime}$).
Let us see the implications of this identification.

In the Lorentzian signature case identifying $\tilde n_i$ with $n_i$ implies
\be
n_i =  \tilde n_i = r_i + {N+1 \over 2} - i - {B \over  N}~.\label{surplustoYoung}
\ee
A first consistency check of   (\ref{surplustoYoung}) is that the numbers on both sides are required to be distinct. Another check is that summing both left hand and right hand sides over $i$ gives a vanishing result. However,
the left hand side is always an integer or half integer, while the right hand side is not unless the number of boxes is a multiple of $N$. In addition,
(\ref{constrn}) gives a constraint on the $r_i$ which is not obeyed by a generic representation of $SU(N)$. In particular it implies that conical surpluses are states with vanishing odd spin $s$ charge.
Thus, after the analytic continuation in $k$, the smooth Lorentzian configurations account for only a part of the light primaries in the ${\cal W}_N$ minimal models.

It turns out that the Euclidean version provides a much more precise matching.
The identification is now
\be
n_i^{\prime} = m_i - {1\over N}\sum_{i} m_i = \tilde n_i = r_i + {N+1 \over 2} - i - {B \over  N}~.\label{surplustoYoung2}
\ee
Both sides again obey the first two consistency checks mentioned in the previous paragraph.
But now we see that the integrality properties are the same on both sides. From
(\ref{surplustoYoung2}), we can consistently identify
\be
m_i=r_i+{N+1 \over 2} - i + l  \, \, \, \, \, ({\rm odd} \,  N)~,
\ee
\be
m_i=r_i+{N+1 \over 2} - i + l -{1\over 2}  \,\,\,\,\, ({\rm even}  \,  N)~,
\ee
where $l$ is an integer chosen so that $r_N=0$.
There is also no longer a reality constraint on $m_i$, which previously restricted $n_i$ to obey (\ref{constrn}). Therefore, we generically will have non-zero odd spin charges.

With this identification we therefore can precisely match the $(\Lambda, \Lambda)$ primaries (analytically continued) of the ${\cal W}_N$ minimal models with the smooth configurations of the $SL(N, \CC)$ theory. The very non-trivial agreement of the spin 2, 3 and 4 charges is compelling evidence for the correctness of the identification.

\section{Discussion}\label{sec:concl}

We have systematically analyzed the spectrum of admissible configurations in $SL(N, \RR)\times SL(N, \RR)$ higher spin theory whose topologies are defined by a contractible spatial cycle. Such solutions are characterized  by their holonomy being a trivial element of the gauge group, which provides a gauge invariant criterion for smoothness.  They form a discrete spectrum which is parametrized by $\lfloor N/2 \rfloor$ integers.

The novelty of these new states is two fold. First,  in the appropriate gauge, the metric is locally AdS with a conical defect at the origin and a specific value of the deficit angle.
Using the higher spin gauge invariance, this can be transformed into a metric which describes a wormhole without a conical deficit.
This illustrates how certain conical singularities (for discrete values of deficit angles) can be resolved in
higher spin theories. This is, in some ways, analogous to how orbifold singularities are resolved in perturbative string theory.

 Secondly, because of these additional states in the spectrum, the mass gap of the theory is affected. For a pure gravitational theory, the gap is  $\Delta=c/24$ which is determined by the energy difference between global AdS and the massless BTZ black hole. In a higher spin theory the gap is now governed by the lowest energy conical deficit relative to global AdS, which is of order
\be
\Delta \sim {c\over 12\epsilon_N} ~.
\ee
Even for $c \sim N$, in the large $N$ limit we have  $\Delta \sim{ N^{-2}}$ due to \eqref{b:bc}. The presence of unusually light ``solitonic'' excitations is a novel feature of these higher spin theories. It is clearly important to understand the implications of this for an effective field theory description of the bulk beyond the classical limit. Note that this feature exists independent of any conjectural duality to a two dimensional CFT.

We have also argued that these solutions must play a non-trivial role in the duality conjectured in \cite{Gaberdiel:2010pz}.  In particular, we found a one to one correspondence with the light primaries of the ${\cal W}_N$ CFT. Our procedure for this matching involves several distinct components. Firstly, we assumed that in the bulk description, in addition to the smoothness condition, the physical states relevant for the duality are only those obeying the Drinfeld-Sokolov condition (\ref{asAdS}) \cite{Campoleoni:2010zq}. Secondly, motivated by the generalized level-rank duality, we proposed that the analytic continuation $k = -(N+1)$ relates the spectrum of light primaries of the ${\cal W}_N$ minimal models to states in the bulk $SL(N)$ higher spin theories. Thirdly, the one to one matching of states is complete for the Euclidean $SL(N, \CC)$ theory rather than the Lorentzian version. The Lorentzian theory captures only a subset of all the light states of the CFT with our definitions of smoothness.

It is clearly important to examine each of these features in more detail. In particular, the novel feature that the Euclidean theory has more physical states compared to the Lorentzian theory demands a better understanding. It can potentially shed some light on the poorly understood role of a Wick rotation in quantum gravity. Perhaps the considerations in
\cite{Witten:2010cx, Harlow:2011ny} might be of help.

The analytic continuation $k=-(N+1)$ is also quite mysterious, but can perhaps be clarified by a study on the CFT side. Its implication on the bulk side (which we have used here) is also
interesting in that it relates the relatively unfamiliar $hs[\lambda]$ higher spin theory with the $SL(N)$ Chern-Simons theory.
One may also view the above continuation as predicting a specific class of smooth geometries in the $hs[\lambda]$ bulk theory. This would be interesting to verify/construct directly.

There are several other interesting questions thrown up by the results here. For instance, three point functions in the CFT between the light primaries and other perturbative states
have been computed in \cite{Papadodimas:2011pf}. It would be nice to see if these can be reproduced by some appropriate scattering calculation off the smooth geometries studied here. This would also require taking into account the scalar fields in the bulk
which has played no role thus far. In the $\lambda=0$ limit (for any finite $N$), the light primaries have recently been understood in terms of twisted sectors in a continuous orbifold of a theory of $N-1$ free bosons \cite{Matthias-orb}. Can one view our smooth configurations as coming from some kind of twisted sector in the bulk? Their description in terms of different holonomy eigenvalues holds out the prospect of making a direct match with the CFT.

\section*{Acknowledgements}
We would like to thank M. Ammon, D. Francia,  M. Gaberdiel, P. Kraus, E. Perlmutter, J. Lapan, A. Maloney, S. Raju and A. Sen for discussions and remarks. We are grateful to the Centro de Ciencias de Benasque Pedro Pascual for hospitality as well as the opportunity to initiate this project. The work of A.C. is supported in part by the National Science and Engineering Research Council of Canada, and in part by the National Science Foundation under Grant No. NSF PHY-05-51164. R.G.'s research is partially supported by a Swarnajayanthi Fellowship of the DST, Govt. of India and more generally by the support for basic sciences from the people of India. The work of MG   was supported in part by NSF grant PHY-07-57702. The work of JR has been supported  in part by the Czech Science Foundation  grant GACR P203/11/1388 and in part by the EURYI grant GACR  EYI/07/E010 from EUROHORC and ESF.

\appendix
\section{Conventions}

\subsection{$SL(N,\RR)$ algebra and generators}\label{app:slN}

In the principal embedding of $SL(2,\RR)$ in $SL(N,\RR)$ we have besides a spin 2 field, fields with spin $s=3,\cdots, N$. $\{L_0, L_{\pm1}\}$ label the $SL(2,\RR)$ subalgebra, and $W_m^{(s)}$  are the spin $s$ generators where $m=-(s-1)\ldots (s-1)$. In this representation we have
 \bea
[L_i ,L_j] &=& (i-j)L_{i+j}~, \cr
[L_i, W_m^{(s)}] &=& (i(s-1)-m)W_{i+m}^{(s)}~.\label{sl2comms}
  \eea

In the $N$-dimensional representation of $SL(N,\RR)$, the $L_i$ generators for the principal embedding of $SL(2,\RR)$ can be taken as
\be\label{ellone}
L_{1}=
- \left(
\begin{array}{ccccccc}
0& \cdots  & & & & &0\\
 \sqrt{N-1}  &  0 & & & \cdots & &  \\
 0 & \sqrt{2(N-2)}  &0 & & & &  \\
 \vdots &                   &\ddots &\ddots&&&\\
  &     & &\sqrt{|i(N-i)|}& 0& &\\
   & &  &  &\ddots & \ddots  &\\
0& \ldots & && & \sqrt{(N-1)}&0
\end{array}
\right)~,
\ee
\be\label{ellminus}
L_{-1}=
\left(
\begin{array}{ccccccc}
 0  &  \sqrt{N-1} & & \cdots &  && 0  \\
 \vdots & 0  &\sqrt{2(N-2)} & & & &  \\
   & \vdots  &\ddots &\ddots & & &  \\
  &   & & 0& \sqrt{|i(N-i)|}& &\\
   &  & &  &\ddots & \ddots  &\\
& & & & & 0&\sqrt{(N-1)}\\
0& \cdots  &  & &  &&0
\end{array}
\right)~,
\ee
and
\be\label{app:L0}
L_0=
{1\over 2} \left(
\begin{array}{ccccccc}
     (N-1)& 0&\cdots  && &  0 \\
  0 &   (N-3) & &&  \vdots\\
  \vdots &  &    (N+1-2i)& &\\
   &  &   &  \ddots& &\\
& & & &-(N-3)&\\
0 &\cdots&  & &&-(N-1)
\end{array}
\right)~.
\ee
From this we find the normalization
\be
\epsilon_N =\tr(L_0 L_0) = {1\over 12} N(N^2-1)~.
\ee
Note that our generators satisfy the hermiticity property
\be
(L_j)^\dagger = (-1)^j L_{-j}~.\label{herm}
\ee

An explicit representation for the other $SL(N,\RR)$ generators is as follows:
\be\label{wsform}
W^{(s)}_m = (-1)^{s-m-1}{(s + m -1)! \over (2 s - 2)!} \underbrace{[L_{-1},[L_{-1},\ldots,[ L_{-1}}_{s-m-1 \,{\rm terms}} ,L_1^{s-1}]\ldots ] ]~.
\ee
They satisfy the hermiticity property
\be
\left( W^{(s)}_m \right)^\dagger = (-1)^m  W^{(s)}_{-m}  ~.
\ee
The Cartan-Killing form on $SL(N,\RR)$ is given by
\bea
{\rm Tr }  W^{(s)}_m W^{(r)}_n = t^{(s)}_m \d^{r,s}\d_{m,-n}~,
\eea
and
\bea\label{app:tr}
t^{(s)}_m = (-1)^m { (s-1)!^2 (s + m -1)! (s-m-1)! \over ( 2s -1)!(2 s-2)!} N \prod_{i = 1}^{s-1} (N^2 - i^2)~.
\eea

\subsection{Generators in the $l$-th off diagonal }\label{diagonalgen}
In order to construct the gauge connections with smooth holonomies it is useful to construct the following generators which have nonzero elements in the $l$-th diagonal above and below of  the main diagonal of the matrix.
\be\label{bblock}
\Big[B^{(l)}_k(x,y)\Big]_{ij} = x \; \delta_{i,k}\delta_{j,k+l}- y \; \delta_{i,k+l}\delta_{j,k}~,
\ee
where $k=1,2,\cdots N-l$. The condition that two generators $B^{(l)}_k$ and $B^{(l)}_{k'}$ commute  is given by
\be
[B^{(l)}_k,B^{(l)}_{k'}]=0  \;\; \Longleftrightarrow \;\;\;  k\neq k'+l, \;\; k\neq k'-l~.
\ee
For example, the case $l=1$ discussed in section \ref{smoothdef} implies that the set of $B^{(1)}_k$  with odd $k$ is a commuting set.

A useful formula to calculate the metric  from the gauge connection in section \ref{smoothdef}
\be
e^{\rho L_0} B^{(l)}_k (x,y) e^{-\rho L_0} = B^{(l)}_k (e^{+l\rho} x, e^{-l \rho} y)~.
\ee

\subsection{Non principal embeddings}\label{nonprinc}
It was shown in \cite{Dynkin:1957um} that inequivalent embeddings (i.e. embeddings which are not related by conjugation) of $SL(2,\RR)$ in $SL(N,\RR)$ are in one-to-one correspondence with integer partitions $P$ of $N$.
\be
P: \qquad N = \sum_{A=1}^N P_A A~.
\ee
The  partition determines a branching of the fundamental representation $N$ under the $SL(2,RR)$, given by
\be
\underline{N} \to \bigoplus_{A = 1}^N P_A \underline{A}~.
\ee
For example, for the principal embedding one has $P_A = (0,0, \ldots, 0,1)$.
The $SL(2,\RR)$ generators in the embedding  $P$ can be taken to be
\be\label{slbloack}
 L_{0,\pm1}^{(P)} = \bigoplus_{A=1}^N \bigoplus_{a = 1}^{P_A} L_{0,\pm1}^{(A)}~,
 \ee
 where $L^{(A)}_{0,\pm1}$  can be obtained from  (\ref{ellone})-(\ref{app:L0}) by replacing  $N$  by $A$.
 The normalization of the trace  for the  embedding  $P$ is
\be\label{epspdef}
\e_P =  \tr (L_0^{(P)})^2= {1\over 12} \sum_A P_A A(A^2-1)~.
\ee
For each $A$ we can define off diagonal generators  $B^{(l,A)}_k$ as in (\ref{bblock}) with the indices restricted  to the   $A\times A$ block.

The most  general matrix with non zero indices directly above and below the diagonal in each $A\times A$ block is given by a direct sum of $B^{(1,A)}_k$
and satisfies
\be
e^{\rho L^{(A)}_0} B^{(1,A)}_k (x,y) e^{-\rho L^{(A)}_0} = B^{(1,A)}_k (e^{+\rho} x, e^{-\rho} y)~.
\ee

\section{Normalizations of higher spin charges}\label{appnorms}
In this Appendix we will verify explicitly that our normalizations of the higher spin generators on the Chern-Simons side (\ref{hwgauge})
and on the CFT side (\ref{ucharges},\ref{chargesCFT}) for $N=3$ are consistent. We do this by comparing the Poisson brackets on the gravity
side with the commutators on the CFT side.

On the Chern-Simons side, the higher spin mode expansion in our conventions is
\be
a_+ = L_1 + {1 \over k_{\rm cs}} \sum_{s=2}^N {k_{\rm cs}^{-{s-2\over 2}}\over t^{(s)}_{s-1}} \left( \sum_m   w^{(s)}_m e^{-im x^+} \right) W_{-(s-1)}^{(s)}~,
\ee
where the $w^{(2)}_m$ are related to the Virasoro modes by $w^{(2)}_m = l_m - {c \over 24 }\d_{m,0}$.
The Poisson brackets between the modes $w^{(s)}_m$ can be computed using the method explained in e.g. \cite{Gaberdiel:2011wb}.
For $N=3$, this gives\footnote{Note that the dictionary between the conventions of \cite{Gaberdiel:2011wb} and our conventions is $k_{\rm cs}^{there} = 4 k_{\rm cs}^{here},\ \s^{there} = -1/4,\ \mathcal{L}_m^{there} = l_m^{here},\ \W_m^{there} = -w^{(3)\, here}_m / \sqrt{k_{\rm cs}}$.}
\bea
i \left\{\, l_m , l_n \,\right\}_{PB} \, &=& (m-n) l_{m+n} + {c \over12} (m^3 - m) \d_{m+n,0} ~,\label{classPoiss1}\\
i \left\{\, l_m , w_n^{(3)} \,\right\}_{PB} \, &=& (2 m  -n) w_{m+n}^{(3)} ~,\label{classPoiss2}\\
 i \left\{\, w^{(3)}_m , w^{(3)}_n \,\right\}_{PB} \, &=& k_{\rm cs} \,\bigg[\, {1\over 12} (m-n)(2m^2+2n^2-mn-8)\, l_{m+n} \, + \,
\frac{8}{c}\, (m-n)\, \l_{m+n}\nonu
&&  + \, \frac{c}{144}\, m(m^2-1)(m^2-4)\, \d_{m+n,0} \,\bigg] \, ,\label{PoissW3}
\eea
where
\begin{equation}
\l_m \,\equiv\, \sum_{n}\, l_{n} l_{m-n}\ .
\end{equation}

On the CFT side, the charges  \eqref{ucharges} and \eqref{chargesCFT} correspond to the following normalization
for the $\W_3$ algebra (see  pg. 237 of \cite{Bouwknegt:1992wg})
\bea
\,[ L_m , L_n ]  &=& (m-n) L_{m+n} + {c \over 12} (m^3 - m) \d_{m+n,0}~,\label{qcomm1}\\
\, [ L_m , W^{(3)}_n ]  &=& (2 m  -n) W_{m+n}^{(3)}~, \label{qcomm2}\\
\,[ W^{(3)}_m , W^{(3)}_n ] &=& - ( \a_0^2 - 4/15)  \,\bigg[\, {1\over 12} (m-n)(2m^2+2n^2-mn-8)\, L_{m+n} \,\cr &&+ \, \frac{8}{c+ 22/5}\, (m-n)\,
\L_{m+n}  + \, \frac{c}{144}\, m(m^2-1)(m^2-4)\, \d_{m+n,0} \,\bigg] \, ,\label{commW3}
\eea
with
\be
\L_m \equiv \sum_n  L_n L_{m-n}  -{3\over 10}(m+3)(m+2) L_m~.
\ee
Using the prescription (\ref{CAC}) to relate the Chern-Simons and CFT computions, and allowing for the usual shifts of the level between classical and
quantum algebras, we see that the normalizations in (\ref{PoissW3}) and (\ref{commW3}) are indeed consistent.

\end{document}